# Equilibrium Molecular Dynamics Study of Lattice Thermal Conductivity/Conductance of Au-SAM-Au Junctions

**Tengfei Luo**  
2555 Engineering Building  
Mechanical Engineering  
Michigan State University  
East Lansing, MI 48824  
Email: luotengf@msu.edu  

**John R. Lloyd***  
ASFC 1.316.C RRMC  
Rapid Response Manufacturing Center  
The University of Texas Pan American  
Edinburg, TX 78539  
Email: lloyd@egr.msu.edu  

* corresponding author

**Abstract**

In this paper, equilibrium molecular dynamics simulations were performed on Au-SAM (self-assembly monolayer)-Au junctions. The SAM consisted of alkanedithiol ($-S-(CH_2)_n-S-$) molecules. The out-of-plane (z-direction) thermal conductance and in-plane (x- and y-direction) thermal conductivities were calculated. Simulation finite size effect, gold substrate thickness effect, temperature effect, normal pressure effect, molecule chain length effect and molecule coverage effect on thermal conductivity/conductance were studied. Vibration power spectra of gold atoms in the substrate and sulfur atoms in the SAM were calculated and vibration coupling of these two parts was analyzed. The calculated thermal conductance values of Au-SAM-Au junctions are in the range of experimental data on metal-nonmetal junctions. The temperature dependence of thermal conductance has similar trend to experimental observations. It is concluded that the Au-SAM interface resistance dominates thermal energy transport across the junction, while the substrate is the dominant media in which in-plane thermal energy transport happens.

**Nomenclature**

| | |
|---|---|
| $\phi$ | potential energy |
| $\vec{F}$ | force vector |
| $m$ | atomic mass |
| $\vec{r}$ | position vector |
| $t$ | time |
| $k$ | thermal conductivity |
| $J(\tau)$ | heat current at instant $\tau$ |
| $V$ | volume of simulation system |
| $k_B$ | Boltzmann constant |
| $T_m$ | mean temperature of a simulation |
| $E_j$ | total energy of atom $j$ |
| $U_s$ | potential energy of bond stretching potential |
| $k_s, r_o$ | parameters for bond stretching potential |
| $U_\theta$ | potential energy of bond bending potential |
| $k_\theta, \theta_o$ | parameters for bond bending potential |
| $U_t$ | potential energy of torsion potential |
| $\varphi$ | dihedral angle of four body group |
| $a_0, a_1, a_2, a_3, a_4, a_5$ | parameters for torsion potential |
| $U_{L-J}$ | potential energy of Lennard-Jones potential |
| $\varepsilon, \sigma$ | parameters for Lennard-Jones potential |
| $U_M$ | potential energy of Morse potential |
| $De, a, r_{mo}$ | parameters for Morse potential |
| $G$ | thermal conductance |
| $q$ | heat flux normal to the junction interfaces |
| $\Delta T$ | temperature difference across the junction |
| $L$ | junction thickness |
| $R$ | thermal resistance |

## 1. Introduction

Molecular electronic devices have become more and more important nowadays, and among them, the self-assembly monolayer (SAM) on metal or semiconductor substrates has drawn much attention. There have been intensive works focused on the electronic and structural properties of SAM-solid junctions [1-4]. However, the studies of thermal properties of such junctions are limited, and knowledge of thermal transport in these junctions is very important to the growing fields of molecular electronics and small molecule organic thin film transistors. Ge et. al [5] measured the transport of thermally excited vibrational energy across planar interfaces between water and solids that have been chemically functionalized with SAM using the time-domain thermoreflectance. Wang et. al [6] studied heat transport through SAM of long-chain hydrocarbon molecules anchored to a gold substrate by ultrafast heating of the gold. Patel et. al [7] studied interfacial thermal resistance of water-surfactant-hexane systems by non-equilibrium molecular dynamics.

There are two typical kinds of SAM-solid junctions, one of them is the SAM-metal junction, which has been studied a lot [8,9]. The other kind is the SAM-semiconductor junction, which is relatively new, but has become very popular [10]. In this work, thermal transport in Au-SAM-Au junctions with alkanedithiols ($-S-(CH_2)_n-S-$) being the SAM molecules is studied using molecular dynamics (MD). The reason such junctions are chosen is that the structural properties, including the absorption site, tilt angle, coverage and etc., have been studied thoroughly [8,9,11,12], and a set of reliable classical potentials for MD simulation is available [13]. In such metal-SAM-metal junctions, the SAM-metal interfaces play important roles in thermal energy transport across the junctions, especially when the system sizes are in the nanoscale [14].

In solid materials, there exist two kinds of heat carriers in thermal energy transport: phonons and electrons. Phonons are the quanta of the lattice vibrational field [15]. Phonons dominate the thermal energy transport in semiconductors and insulators while electrons play important roles in energy transport in metals. In this work, where the metal-SAM junction exists, it is difficult for electrons in the metal to tunnel through the SAM molecules, and thus electron transport is largely depressed by the nonmetal SAM layer. As a result, this work focuses on the phonon part of thermal transport and calculates the lattice thermal conductivities/conductance in both in-plane directions (x- and y-directions) and out-of-plane direction (z-direction) of the junctions.

For lattice thermal conductivity calculations, classical MD with appropriate potential functions has been demonstrated to be a powerful method [16-20]. However, to our knowledge, no work has been done to investigate the thermal transport properties of metal-SAM-metal junctions by MD simulation. In this work, equilibrium classical MD simulations on Au-SAM-Au junctions are performed and thermal transport properties are calculated using Green-Kubo method [22]. Thermal conductivities and conductance are calculated versus simulation cell size, temperature, junction thickness, molecular chain length, molecule coverage (number of molecules on Au substrates) and simulated normal pressure.

## 2. Theory and Simulation

Classical MD is a computational method that simulates the behavior of a group of atoms by simultaneously solving Newton's second law of motion (eq.(1)) for the atoms with a given set of potentials.

$$-\nabla \phi = \vec{F} = m\frac{d^2\vec{r}}{dt^2}. \qquad (1)$$

where $\phi$ is the potential energy, $\vec{F}$ is force, $m$ is atomic mass, $\vec{r}$ is position vector and $t$ is time. By processing the trajectory obtained from an MD simulation using different statistical techniques, transport properties such as diffusion coefficient and thermal conductivity can be calculated.

There are two different MD approaches to study thermal transport in solid systems: equilibrium method and non-equilibrium method (also called direct method). The non-equilibrium method is like an experiment. By applying a heat bath to a system, the thermal conductivity is calculated by Fourier's law of conduction [7,21]. In the equilibrium method, a system is simulated in an equilibrium state. It relies on small statistical temperature fluctuations to drive instantaneous heat fluxes. The thermal conductivity of the system can be calculated from the heat current autocorrelation (HCAC) function according to the Green-Kubo formula [22]. This method has been used widely and has been shown to give good results [23,24].

According to Green-Kubo relation, the thermal conductivity tensor $k$ can be calculated with the following equation:

$$k = \frac{1}{V k_B T_m^2} \int_0^\infty \langle \vec{J}(\tau) \vec{J}(0) \rangle d\tau \qquad (2)$$

$\vec{J}(\tau)$ is the instantaneous heat current and $\langle \vec{J}(\tau) \vec{J}(0) \rangle$ is the heat current auto-correlation function. $V$ is the volume of the simulated system. $k_B$ is Boltzmann constant. $T_m$ is the mean temperature during the production period of a MD simulation. Heat current $\vec{J}$ is given by:

$$\vec{J} = \frac{\partial}{\partial t}\left(\sum_j \vec{r}_j E_j\right) \qquad (3)$$

where the subscript $j$ refers to the index of atoms. $E_j$ is the total energy including potential and kinetic energy of atom $j$. Since eq. (3) produces heat current vectors, thermal conductivities in x-, y-, z-directions can then be calculated by one single simulation.

It needs to be noted that in this work, the calculated thermal conductivity in z-direction is a combination of thermal conductivities of the materials making up the junction. It should not be considered as a material property and its value varies when the system configuration changes.

In any MD simulation, potential functions are critical. In this work, the well established Hautman-Klein model [13] is employed. In this model, the light hydrogen atoms of the hydrocarbon molecules are not simulated explicitly but incorporated into the carbon backbone. Their masses are added to the carbon atoms which they bond to, and forming pseudo carbon atoms. The reason for such treatment is that the high frequency vibrational motions of light hydrogen atoms in the hydrocarbon groups are less important than the lower frequency movements of the carbon backbone [13]. Such a treatment has been demonstrated to be a valid approach to simplify simulations and to give good results [24,25]. Bond stretching, bond bending and Ryckaert-Bellemans torsion potentials are used in alkanedithiol molecules for the bonded interactions. Morse type interactions are used to simulate the interaction between sulfur atoms and gold substrates [26,27] and the interaction among gold atoms [28]. The Lennard-Jones potentials together with the Lorentz-Berthelot mixing rule [29], $\varepsilon_{ab} = \sqrt{\varepsilon_a \varepsilon_b}$, $\sigma_{ab} = \frac{1}{2}(\sigma_a + \sigma_b)$, are used to simulate long distance and intermolecular interactions. The potential function and parameters are listed in Table 1. All the parameters are from ref. [13] and [26], or as otherwise specified in the table. In our simulations, a neighborlist with cutoff of less than 12.00 Å is used to speed up the calculation (In our code, the neighborlist cutoff radius is compared with dimensions of the system, and it is adjusted automatically to avoid double counting). This neighborlist is not updated after the structure is optimized due to the fact that there should not be large atomic displacements other than vibrations about the equilibrium positions in the solid phase system. For every simulation, 5 separate runs with different random initial conditions were performed. The resulting values are averaged over the 5 runs.

| Potential | Function forms and parameters |
|---|---|
| Bond Stretching<br>$S-C$<br>$C-C$ | $U_s = \frac{1}{2}k_s(r-r_o)^2$           (4)<br>where<br>$k_s = 14.00224 eV / \text{Å}^2$<br>$r_o = 1.523 \text{Å}$ (for $C-C$); $1.815 \text{Å}$ (for $S-C$) |
| Bond Bending<br>$S-C-C$<br>$C-C-C$ | $U_\theta = \frac{1}{2}k_\theta(\theta-\theta_o)^2$           (5)<br>where<br>$k_\theta = 5.388 eV / rad^2$<br>$\theta_o = 109.5$ (for $C-C-C$); $114.4$ (for $S-C-C$) |
| Ryckaert-Bellemans Torsion<br>$S-C-C-C$<br>$C-C-C-C$ | $U_t = \sum_{i=0}^{5} a_i \cos^i(\varphi)$           (6)<br>where<br>$\varphi$ is the dihedral angle<br>$a_0 = 0.09617 eV, a_1 = 0.125988 eV, a_2 = -0.13598 eV$<br>$a_3 = -0.0317 eV, a_4 = 0.27196 eV, a_5 = -0.32642 eV$ |

| | | |
|---|---|---|
| Lennard-Jones with Lorentz-Berthelot mixing rule | $U_{L-J} = 4\varepsilon \left[ \left(\dfrac{\sigma}{r}\right)^{12} - \left(\dfrac{\sigma}{r}\right)^6 \right]$ <br><br> where $\varepsilon$ and $\sigma$ are determined by mixing rule [29] <br><br> for $C$: $\varepsilon = 0.00513 eV, \sigma = 3.914 \text{ Å}$ <br><br> for $S$ that interact with other atoms other than $S$: <br><br> $\varepsilon = 0.01086 eV, \sigma = 3.550 \text{ Å}$ <br><br> for $Au$: $\varepsilon = 0.001691 eV, \sigma = 2.934 \text{ Å}$ [30] <br><br> for $S-S$ interaction: $\varepsilon = 0.01724 eV, \sigma = 4.250 \text{ Å}$ | (7) |
| Morse <br> $Au-Au$ <br> $Au-S$ | $U_M = De \left[ \left(1 - e^{-a(r-r_{mo})}\right)^{12} - 1.0 \right]$ <br><br> where <br> $De = 0.475 eV$ (for $Au-Au$)[39]; $0.380 eV$ (for $Au-S$)[28] <br> $a = 1.583 \text{ Å}^{-1}$ (for $Au-Au$); $1.470 \text{ Å}^{-1}$ (for $Au-S$) <br> $r_{mo} = 3.0242 \text{ Å}$ (for $Au-Au$); $2.650 \text{ Å}$ (for $Au-S$) | (8) |

Table 1. Potential Functions with Parameters Used in the Simulations

The structures of Au-SAM-Au junctions studied in this work are shown in Figure 1. The junctions consist of two gold substrates with SAM in between. $-S-(CH_2)_8-S-$ is used as the SAM molecule for all simulations except those in section 3.7. Figure 1 shows systems with different cell sizes in x- and y-directions. In this figure, (a) is a system with 4 molecules; (b) is a system with 16 molecules and (c) contains 36 molecules. In all these systems, each substrate contains 12 layers of gold atoms. The procedures to prepare the junctions are as follow: first, one Au(111) substrate is optimized using the Morse potential (Figure 2(a)). Then, a number of SAM molecules are implanted on the gold substrate with the sulfur heads placed in the three-fold hollow sites of the Au(111) surface, thus forming a lattice with lattice constant of 4.99 Å [8,12,31] (Figure 3). The three-fold site was reported to be the most stable site for SAM molecules absorption [8,31]. The molecules are then relaxed at 100K until all molecule atoms attained their equilibrium positions (Figure 2(b)). Finally, the other optimized gold substrate is imposed on top of the SAM (Figure 2(c)). The equilibrium cell dimension in the z-direction was found to be about 68.10 Å. It was reported that the tilt angle does not change much at a temperature range from 50K to 300K [31]. In this work, temperature effect on the tilt angle was ignored so that the simulation cell size in z-direction remains unchanged throughout the temperature range (50K-350K).

In MD simulations, there is a limitation on the number of atoms that can be handled due to the capability of the computing hardware. Thus the thickness of the gold substrate must be limited to tens of angstroms. However, in reality, the gold substrates are often thick (several hundred microns) and their properties are close to bulk solids. The periodic boundary condition (PBC) in z-direction was used to compensate the thickness limit in our simulations. The validation of this treatment is discussed in section 3.2. Besides PBC, free boundary condition is also used in some simulations for comparison. PBCs are used in x- and y- directions with no exception. A simulation system which contains 4 SAM molecules and 12 gold atoms in each substrate layer is defined as a unit cell (see Figure 1(a)). It has a dimension of 9.99 $\overset{o}{A}$ in x-direction and 8.652 $\overset{o}{A}$ in y-direction. The unit cell is expanded in the x- and y- directions to obtain the simulation supercells. Supercells with sizes of 1x1, 2x2 and 3x3 unit cells, which correspond to systems with 328, 1312 and 2952 atoms are presented in Figure 1.

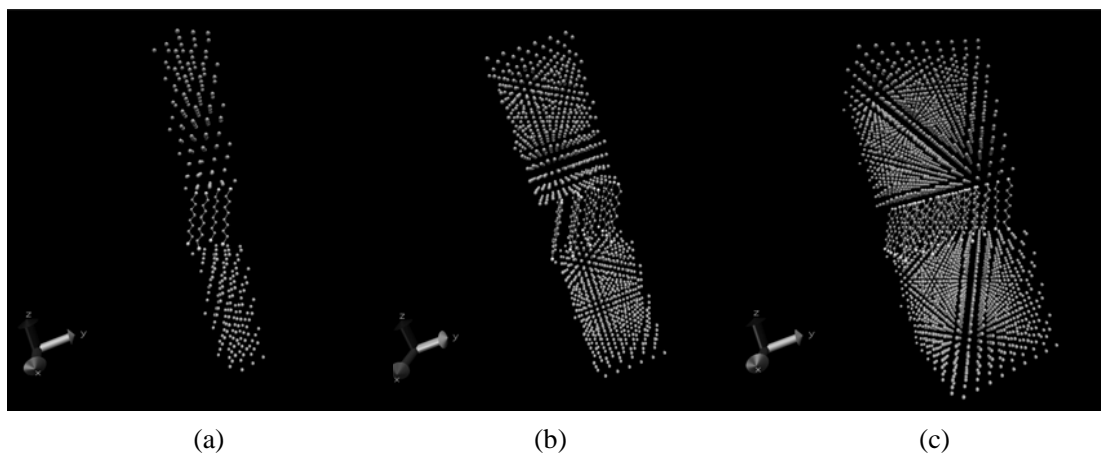

(a)  (b)  (c)

Figure 1. Simulated Au-SAM-Au systems of different sizes: (a). a 1x1 system of 328 atoms (4 alkanedithiol molecules and 288 gold substrate atoms), (b). a 2x2 system of 1312 atoms (16 alkanedithiol molecules and 1152 gold substrate atoms), (c). a 3x3 system of 2952 atoms (36 alkanedithiol molecules and 2592 gold substrate atoms).

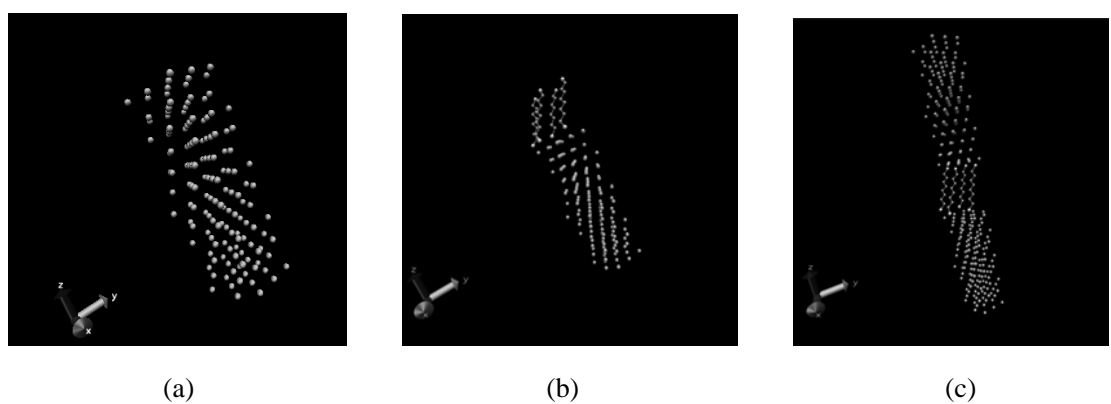

(a)  (b)  (c)

Figure 2. Procedures of preparing the Au-SAM-Au simulation system: (a). One gold substrate is optimized by Morse potential; (b). Alkanedithiol molecules are implanted on the substrate and the whole system is relaxed using the potentials specified in table 1; (c). The other optimized substrate is imposed on top of the alkanedithiol molecules.

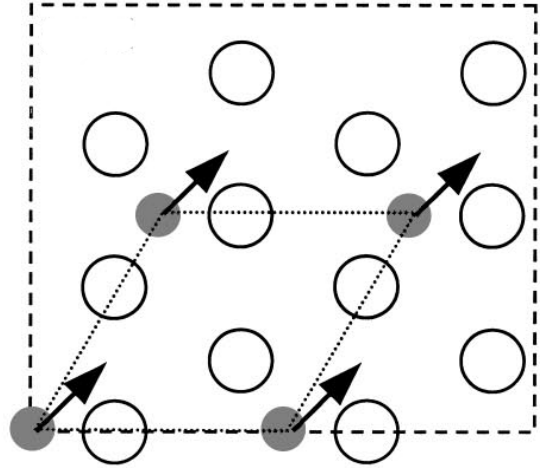

Figure 3. Absorption sites and tilt directions of SAM molecules on Au(111) surface [31] (Open circles are gold atoms on the surface of substrate. Filled circles are sulfur heads which are absorbed on the substrate surface. Arrows represent tilt directions of SAM molecules. Dashed lines form the boundary of the simulation system and dotted lines represent SAM lattice constants.)

The simulation procedure is as follow: (1) All atoms started moving from their equilibrium positions with random initial velocities. (2) Nose-Hoover thermostats were applied to the system for a long time (> 25ps) to make sure that the system reached the target temperature. (3) Thermostats were then released and an equilibration period of 150ps was performed. (4) A production run was performed in which the HCAC function was calculated. (5) HCAC function was integrated and the resulted thermal conductivity was plotted to find a convergence value.

## 3. Results

### 3.1. Defining the Value of Thermal Conductivity

A typical normalized heat current auto-correlation (HCAC) function profile is presented in Figure 4. It can be seen that fast fluctuations are imposed on the overall profile, which make the integration in eq. 2 non-trivial. There are three possible factors that cause these fast fluctuations: (1) high frequency optical phonons; (2) junction interfaces (SAM-Au interfaces) which are only several nanometers away from each other; (3) boundary conditions which result in additional scattering of the phonons. The calculated thermal conductivity profiles are different in different simulations (see Figure 5). For thermal energy transport in the out-of-plane direction with free boundary conditions, phonons traveling across the junction will be scattered at two Au-SAM interfaces, and they will also be scattered at the free Au surfaces at the ends of the simulation cell. Then this will further reduce the out-of-plane thermal conductivity. However, in the in-plane directions, no such scattering exists, and the thermal conductivity exhibits a smoother profile in these directions. To get the thermal conductivity from different profiles, different methods were used to find the convergence areas. For thermal conductivity profile, like in Figure 5(a), where a flat exists, the convergent thermal conductivity was defined by averaging the values over the flat area. Such profiles are encountered in some integrations for z-direction thermal conductivities. For profiles like Figure 5(b), which appeared in many z-direction integrations, a convergence area is chosen where the overall profile fluctuates around a mean value, and the overall fluctuation (overall fluctuation refers to the fluctuation with lower frequency compared to the high frequency fluctuations) magnitude becomes a minimum. Thermal conductivities are found by averaging values over the convergence area. For systems with free boundary conditions in z-direction, the integration profiles are shown to be like Figure 5(c), where the value converges to 0. For profiles seen in Figure 5(d), which are found in most integrations in x-, y-directions, the averaged value around the first overall peak is used. In all the different profiles except 5(c), the data is averaged over no less than 5000 steps, which equals to 2.5 ps.

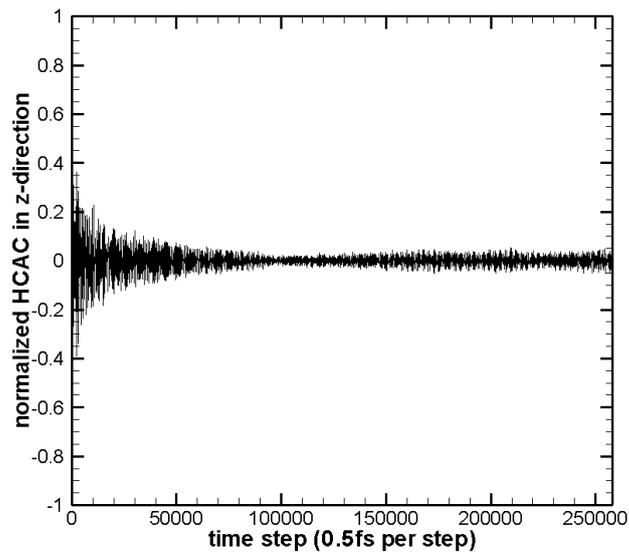

Figure 4. A Typical Normalized Heat Current Auto-Correlation (HCAC) Function

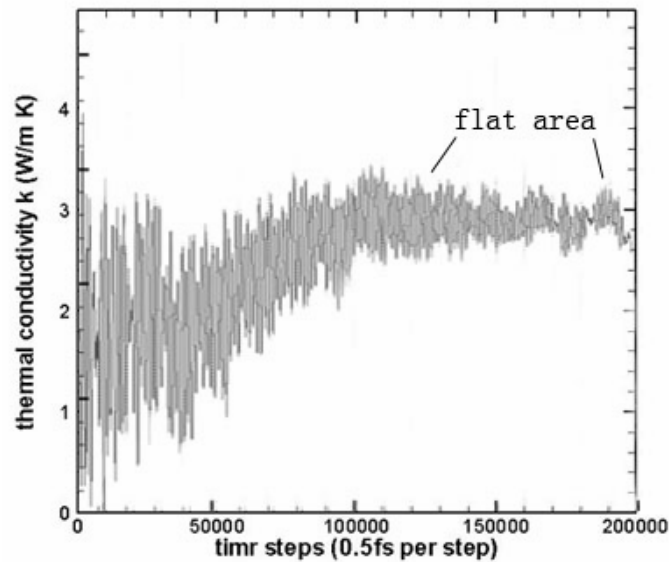

**5 (a)**

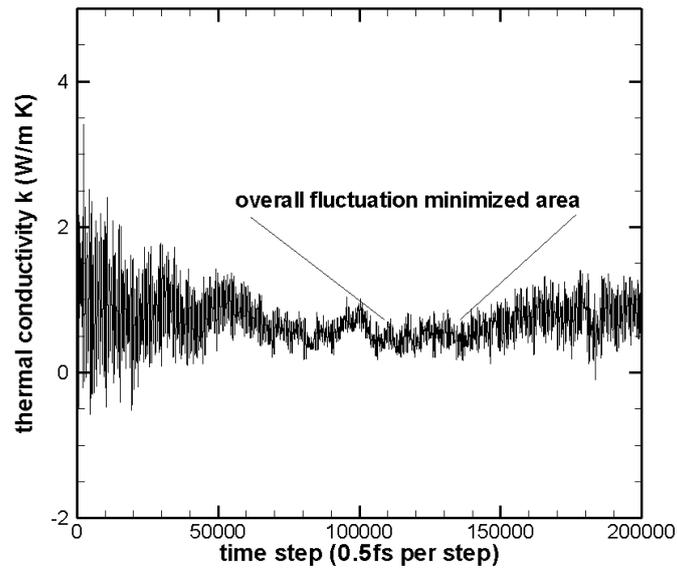

5 (b)

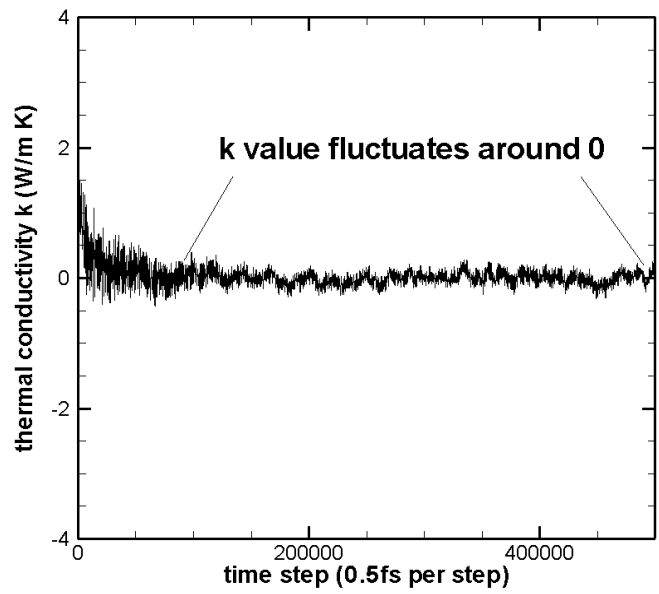

5 (c)

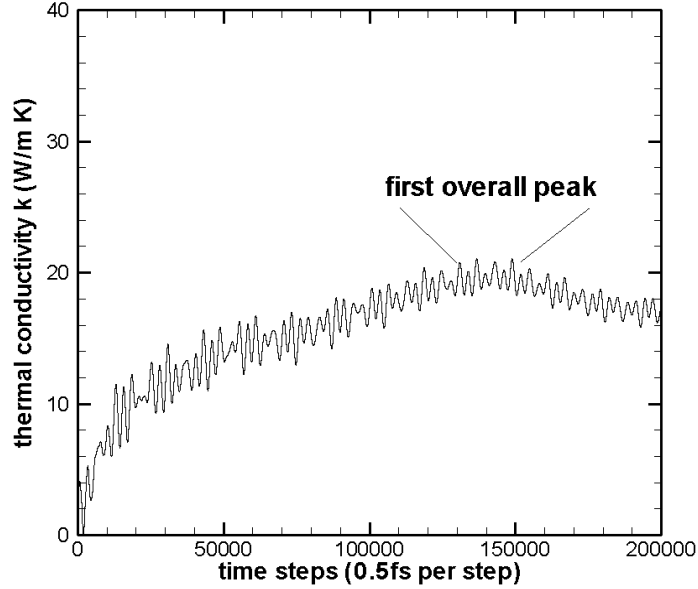

5 (d)

Figure 5. Different thermal conductivity profiles obtained from the calculations: (a). profile that has a flat area; (b). profile that has an overall fluctuation minimized area; (c). profile whose value fluctuates about 0; (d). profile with a first overall peak.

**3.2. Boundary Conditions**

As seen in Figure 5(c), the simulation with the free boundary condition gave thermal conductivities which were approximately 0. There are two possible reasons for such results: (1) the free boundaries add very large boundary resistance to the entire Au-SAM-Au junction which makes the conductivity of the junction very small; or (2) the ultra thin (2-5nm) substrates depressed the excitation of many phonon modes so there were only very limited modes available for thermal energy transport. However, in practice, the substrates are not as thin as several tens of nanometers. To make our simulation reflect greater reality, PBC is used in z-direction. With PBC in z-direction and an appropriate neighborlist cutoff (less than 12 $\text{Å}$ in this work), the gold atoms in one substrate can interact with the image gold atoms of the other substrate while they do not interact with the image SAM molecule atoms if the interaction cutoffs are smaller than the substrate thickness. In this sense, the gold substrates are not isolated thin layers with free surfaces at the junction ends, but rather, they work as thick chunks of gold.

**3.3. Finite Size Effect**

In MD simulations, the sizes of the simulation cells usually affect the results. To investigate the finite size effects in x- and y-directions, thermal conductivity calculations were performed at 100K on 1x1, 2x2 and 3x3 systems with 12 layers of gold atoms on each substrate. Both PBC and free boundary condition were used in the z-direction. PBCs are used in the x- and y-directions for all cases. The results are listed in table 2. The results from 2x2 and 3x3 systems do not differ from each other significantly when the errors are taken into account. This means that, in x- and y-directions, the 2x2 system is sufficient for our simulations. In the rest of the work, all simulations use the 2x2 system.

| System Size | 1x1 | 2x2 | 3x3 | 2x2 | 3x3 |
|---|---|---|---|---|---|
| Boundary condition in z-direction | PBC | PBC | PBC | Free | Free |
| # of atoms | 328 | 1312 | 2952 | 1312 | 2952 |
| $k_x (W/m \cdot K)$ | $4.9 \pm 1.0$ | $17.6 \pm 3.0$ | $22.5 \pm 7.4$ | $12.5 \pm 2.7$ | $13.8 \pm 2.7$ |
| $k_y (W/m \cdot K)$ | $4.6 \pm 1.0$ | $19.4 \pm 3.4$ | $14.6 \pm 3.3$ | $10.8 \pm 1.6$ | $9.5 \pm 1.7$ |
| $k_z (W/m \cdot K)$ | $1.2 \pm 0.2$ | $1.8 \pm 0.3$ | $1.8 \pm 0.4$ | | |

Table 2. Finite Size Effect on Thermal Conductivities

**3.4. Substrate Thickness Effect**

In the simulations, the z-direction is a special direction since it is the cross direction of the Au-SAM-Au junction. Energy transport has to cross two Au-SAM interfaces in this direction. In order to investigate thermal transport across the interfaces, the influence from the limited thicknesses of the substrates should be minimized. The substrate thickness effect on thermal conductivities/conductance for a 2x2 system was studied by changing the number of gold layers in the substrates. All the following cases have PBCs in all three directions and all simulations are carried out at 100K. To investigate the effect of SAM molecules on the substrate in-plane thermal conductivities, the system without SAM molecules is studied. The results are presented in Table 3.

| # of gold layers | 6 | 12 | 18 | 24 | 36 | 12(no SAM) |
|---|---|---|---|---|---|---|
| thickness L of the whole junction ($\mathring{A}$) | 39.42 | 68.10 | 96.73 | 125.67 | 182.90 | 68.10 |
| $k_x (W/m \cdot K)$ | $9.1 \pm 1.5$ | $17.6 \pm 3.0$ | $23.9 \pm 5.6$ | $27.9 \pm 4.3$ | $36.3 \pm 8.3$ | $18.6 \pm 1.6$ |
| $k_y (W/m \cdot K)$ | $7.8 \pm 1.1$ | $19.4 \pm 3.4$ | $23.5 \pm 4.5$ | $19.8 \pm 6.2$ | $25.3 \pm 7.4$ | $19.8 \pm 4.7$ |
| $k_z (W/m \cdot K)$ | $1.3 \pm 0.2$ | $1.8 \pm 0.3$ | $2.5 \pm 0.6$ | $3.8 \pm 1.2$ | $4.9 \pm 0.5$ | 0 |
| $G (MW/m^2 K)$ | $327 \pm 58$ | $261 \pm 46$ | $258 \pm 66$ | $302 \pm 95$ | $268 \pm 30$ | |

Table 3. Substrate Thickness Effect on Thermal Conductivities and Thermal Conductance

To compare the z-direction results, thermal conductance should be the property to be compared, as it is mentioned in section 2 that the thermal conductivity $k_z$ is structural dependent rather than an intrinsic property of the junctions. Thermal conductance, $G$, is defined by $q = G \bullet \Delta T$ where $q$ is the heat flux normal to the junction interfaces, and $\Delta T$ is the temperature difference across the junction. $G$ is related to $k_z$ by

$$G = k_z / L \qquad (9)$$

where $L$ is the thickness of the junction. In Table 3, the thermal conductance does not exhibit monotonic decrease when the substrates are thickened from 6 layers to 36 layers of gold as one would intuitively expect. It is believed that the discrepancies of thermal conductance among systems with different Au thicknesses are not from the thickness size effect but from other factors such as the difficulty of defining the convergence thermal conductivity value and the limited number of runs used to get the mean values. The effects of these factors are reflected in the error bars. The total thermal resistance of the junction can be written as a serial combination of resistances of different parts that make up the junction:

$$R_{total} = 2 \times R_{substrate} + 2 \times R_{interface} + R_{SAM} \qquad (10)$$

Consider the relation between resistance and conductance

$$G = 1/R, \qquad (11)$$

Eq. 10 becomes

$$\frac{1}{G_{total}} = 2 \times \frac{1}{G_{substrate}} + 2 \times \frac{1}{G_{interface}} + \frac{1}{G_{SAM}}. \qquad (12)$$

Due to the large conductance of gold substrates, the first term, $1/G_{substrate}$, becomes negligibly small. (In reality, bulk gold has a thermal conductivity of $318\ W/m \cdot K$, which leads to a conductance $G$ of about $1.2 \times 10^5\ MW/m^2 \cdot K$ for a substrate with thickness of $26.2\ \overset{o}{A}$. The corresponding resistance $R$ is $8.3 \times 10^{-6}\ m^2 \cdot K/MW$ and it is negligible compared to other terms. As stated in section 1, the electron transport contribution to the thermal transport is ignored since it is hard for electrons to tunnel through the SAM molecules. Due to this and the substrate thickness effects, the Au lattice thermal conductivity should be smaller than $318\ W/m \cdot K$. However, even if the calculated thermal conductivities ($\sim 20\ W/m \cdot K$) in the x- and y-directions are used, the resistance comes out to be $\sim 1.3 \times 10^{-4}\ W/m \cdot K$, which contributes only 3% of the total resistance). Wang et. al [6] found that the energy transport along the SAM molecule chains was ultrafast (0.95nm/ps), which suggests that the thermal resistance inside the molecule itself is very small. Therefore, the junction thermal resistance is dominated by the SAM-Au interfaces. Since there are no strong inter-molecular interactions among the discrete SAM molecules, the energy transport from one molecule to another should be weak. Comparing the results of the systems with and without SAM molecules (the 6$^{th}$ case in Table 3), one can see that the existence of SAM molecules does not affect the in-plane thermal transport. It is also found that for the system without SAM molecules connecting the two Au substrates, the out-of-plane thermal conductance becomes 0 as expected. We then believe that the in-plane (x-, y-direction) thermal transport mostly happens in the crystalline gold substrates and SAM molecules present channels for out-of-plane thermal conduction. From Table 3, it can be seen that in the x-, y-directions, the supercell with thickness of 68.10 $\overset{o}{A}$ still suffers from the finite size effect for the in-plane thermal transport. However, if the dimension in z-direction is too large, the calculation becomes too demanding to handle with the current code. For the junctions in our work, what is really important is the out-of-plane (z-direction) thermal transport, and the 12 layer substrates have shown to be thick enough to ignore the thickness effect on the thermal conductance in z-direction (see Table 3). As a result, substrates with 12 gold layers are used in all the rest simulations.

**3.5. Temperature Effect**

One of the topics investigated in this work is the temperature dependence of thermal conductivity/conductance. Figure 6 shows the in-plane results of a system with free boundary conditions. The solid lines are power fits to the discrete data. Data for simulations that use PBC in the z-direction are shown in Figure 7. From the fitted line, the temperature dependence of the calculated lattice thermal conductivities in the in-plane directions (x-, y-directions) shows the same trend as that found in simulations with free-boundary conditions (Figure 6), which declines with increased temperature, and the same trend is found in thermal conductivities of crystalline solids [20,32]. For thermal conductance in the out-of-plane direction (Figure 7(b)), a comparison with Wang et. al [33] experimental data on Au-SAM-GaAs junctions was done (see Figure 10). In Wang's work, Au-alkanedithiols-GaAs junctions were studied which are different from the Au-alkanedithiols-Au junctions studied here. In practical experiments, the alkanedithiols lattice on the GaAs substrate is not as perfect as that on the Au substrate in our work, and the molecules are not always upstraight while some of them lay down. As a result, the effective molecule-substrate contacts are not as good as the ones studied in this paper which are perfect contacts [34]. So, it is not a surprise that our data are much larger than Wang's experimental data. Although the absolute values are not comparable, the trend of the discrete data is similar. The mean thermal conductance increases at low temperatures as the temperatures raises, and becomes almost unchanged at higher temperatures (>150k). We are not aware of any experimental thermal conductance data on exact the same junction that is studied here at a temperature range from 50K to 350K. However, for metal-nonmetal interfaces, the experimental thermal conductance is reported to be $8 < G < 700$ MW/(m$^2$K) [35-37]. In our work, as can be seen in Figure 7(b), the thermal conductance ranges from $200-300$ MW/m$^2$K which falls in the above range. Ge et. al [5] reported Au-(hydrophobic SAM)-water interface to have thermal conductance of $50 \pm 5$ MW/(m$^2$K) and a conductance of $100 \pm 20$ MW/(m$^2$K) for Au-(hydrophilic SAM)-water interface. Wang et. al. [6] reported a thermal conductance of $220 \pm 100$ MW/(m$^2$K) for a Au-SAM junction. It can be seen that our data is on the same order of available experimental data.

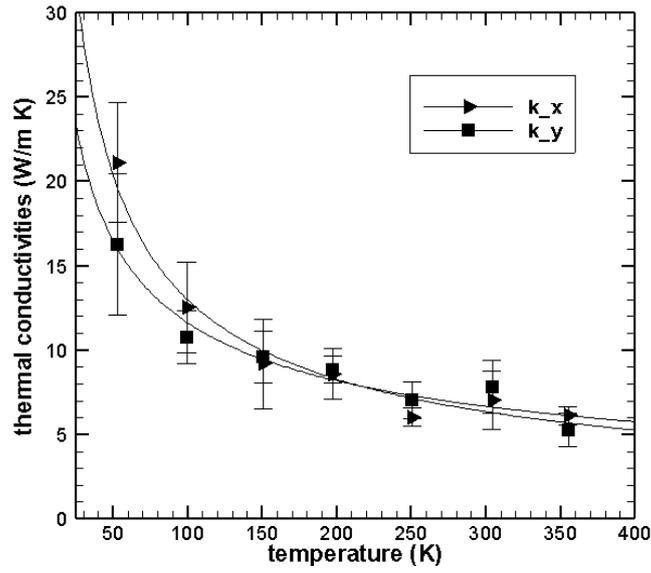

Figure 6. Temperature Dependence of In-plane Thermal Conductivities of Systems with Free Boundary Condition in z-direction

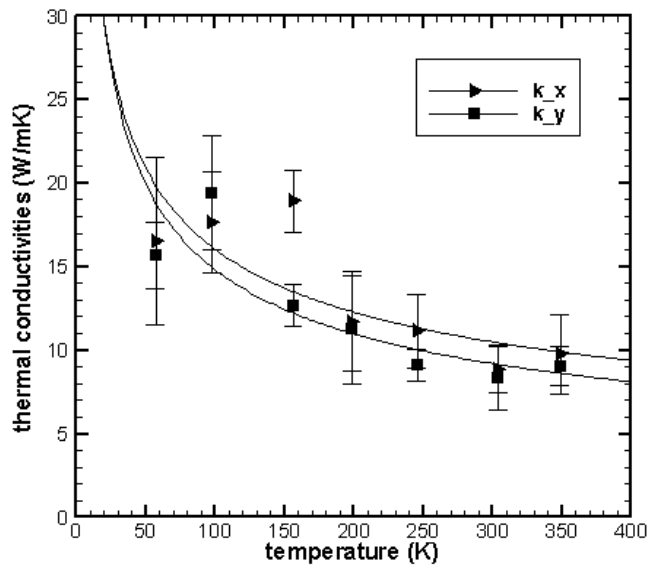

(a) Thermal Conductivities in x-,y-,z-directions

T. Luo and J. R. Lloyd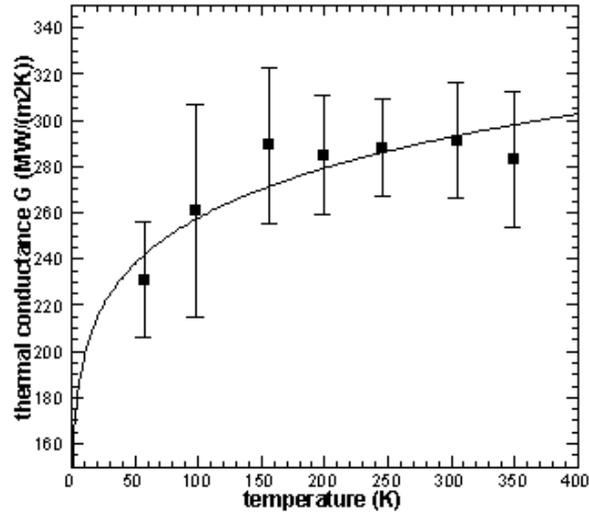

(b) Thermal Conductance of the Au-SAM-Au Junction

Figure 7. Temperature Dependence of Thermal Conductivities and Conductance of Systems with PBC in z-direction

### 3.6. Simulated Normal Pressure Effect

The external normal pressure effect on the thermal conductivity/conductance was simulated by decreasing the dimension of the simulation cell in z-direction at 100K because there is no way to apply forces on the free surfaces of the Au substrates when PBC is used in the z-direction. In this way, the junction would feel "pressure" as the junction is compressed by the decreased cell dimension in z-direction. Results are shown in Table 4 and Figure 8. In Figure 8, where the z-direction thermal conductance presented, no pressure dependence was observed when error bars are considered. The reason for such a result could be that the SAM-Au interface resistances are the dominate factors which impeded the thermal energy transport across the junction (detailed discussions were presented in section 3.4). Although the junction is under pressure, due to the flexibility of the chain-like alkanedithiol molecules, the structure will adapt itself to the small dimension change in z-direction and thus the local dynamics around the SAM-Au interface does not change much. As a result, the interface resistance is not affected much. The thermal conductivities in x- and y- directions do not show any pressure dependence (see Table 4).

| Thickness ($\mathring{A}$) | $k_x (W/m \cdot K)$ | $k_y (W/m \cdot K)$ | $k_z (W/m \cdot K)$ | $G (MW/(m^2 K))$ |
|---|---|---|---|---|
| 68.10 | 22.2 ± 3.4 | 16.4 ± 3.0 | 1.8 ± 0.3 | 265 ± 44 |
| 67.90 | 14.5 ± 4.8 | 21.8 ± 10.2 | 2.0 ± 0.6 | 295 ± 88 |
| 67.70 | 15.9 ± 3.2 | 21.7 ± 6.5 | 2.1 ± 0.4 | 310 ± 59 |
| 67.50 | 14.7 ± 6.6 | 18.1 ± 6.3 | 2.2 ± 0.3 | 326 ± 44 |
| 67.30 | 18.1 ± 4.6 | 17.3 ± 4.0 | 2.3 ± 0.2 | 342 ± 30 |
| 67.10 | 18.3 ± 8.0 | 18.1 ± 7.2 | 2.2 ± 0.2 | 328 ± 30 |

Table 4. Simulated Pressure Dependence of Thermal Conductivities/conductance in x-, y- and z-directions

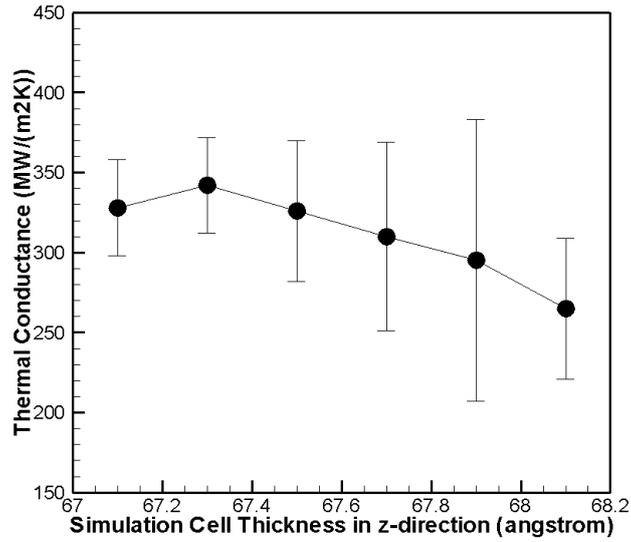

Figure 8. Out-of-plane Thermal Conductance vs. Simulation Cell Thicknesses

### 3.7. Chain Length Effect

Systems with different alkanedithiol molecule chain lengths were also studied. The thermal conductivities of junctions with $-S-(CH_2)_8-S-$, $-S-(CH_2)_9-S-, -S-(CH_2)_{10}-S-$ were calculated at temperatures ranging from 50K to 350K. The in-plane thermal conductivities at 100K are tabulated in Table 5, and they do not exhibit any chain length dependence. The out-of-plane thermal conductance is plotted in Figure 9. Solid lines are power fits of discrete data.

| Chemical Formula | $S_2(CH)_8$ | $S_2(CH)_9$ | $S_2(CH)_{10}$ |
|---|---|---|---|
| $k_x (W/m \cdot K)$ | $17.6 \pm 3.0$ | $14.9 \pm 4.1$ | $22.1 \pm 8.2$ |
| $k_y (W/m \cdot K)$ | $19.4 \pm 3.4$ | $19.3 \pm 3.4$ | $16.3 \pm 1.7$ |

Table 5. Chain Length Effect on Thermal Conductivities in x-, y-directions at 100K

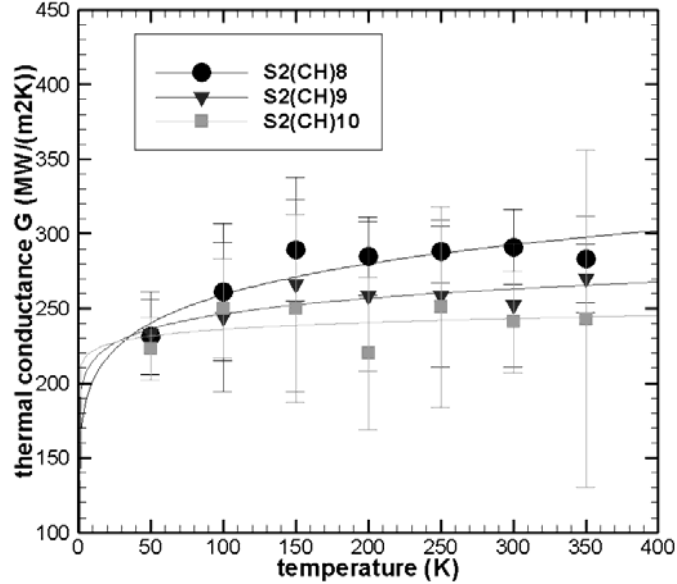

Figure 9. Temperature Dependence of Au-SAM-Au Junction Thermal Conductance with Different Alkanedithiol Chain Lengths

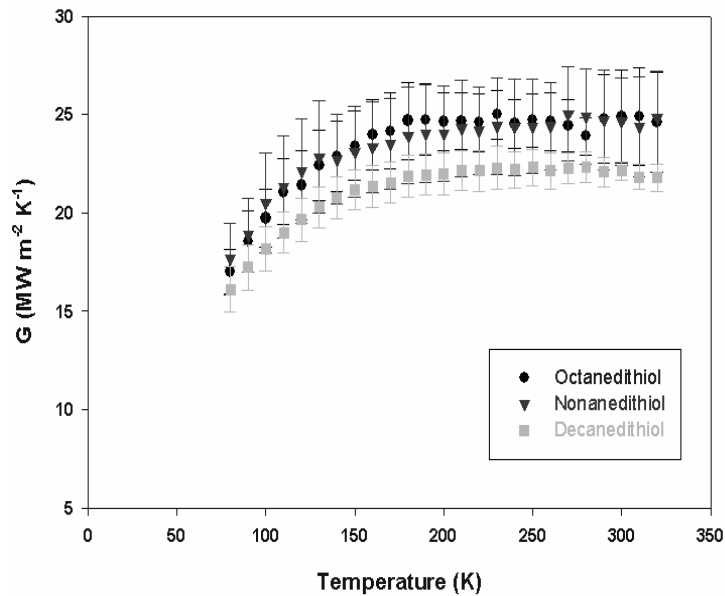

Figure 10. Experimental Data of Temperature Dependence of Au-SAM-GaAs Junction Thermal Conductance with Different Alkanedithiol Chain Lengths [33]

Considering the errors, the thermal conductance values of junctions with different chain lengths are not very different from each other. However, the mean values revealed weak chain-length dependence: as the chain become longer, the out-of-plane thermal conductance decreased slightly. This makes sense if one considers the extreme cases of chain lengths equal to 0 and infinity. The thermal conductance should decrease from the maximum value for junctions with no SAM to 0 when the substrates are separated by infinitely long chains. Such a trend coincides with the experimental measurements of Wang et. al [33] on Au-SAM-GaAs junctions (see Figure 10). The reason that there is no obvious thermal conductance change should still be that the SAM-Au interface resistance dominates the energy transport ability in z-direction while the chain-like molecules themselves have very small resistances. It can be concluded that the limited chain length change (from $-S-(CH_2)_8-S-$ to $-S-(CH_2)_{10}-S-$) does not have significant effect on thermal transport ability of the junctions.

### 3.8. SAM Molecule Coverage Effect

To change the molecule coverage on the Au substrate, the number of molecules attached to the substrate surface was changed. With PBCs in x- and y-directions, every alkanedithiol molecule is equivalent. Molecules were deleted symmetrically so as to keep symmetries. Thermal conductivities/conductance of systems with 16, 14, 12, 10, 8 alkanedithiol molecules were calculated and presented in Table 6.

| Coverage (# of thiols/ simulation cell) | $k_x(W/m \cdot K)$ | $k_y(W/m \cdot K)$ | $k_z(W/m \cdot K)$ | $G(MW/(m^2 K))$ |
|---|---|---|---|---|
| 16 | $17.6 \pm 3.0$ | $19.4 \pm 3.4$ | $1.8 \pm 0.3$ | $264 \pm 44$ |
| 14 | $13.1 \pm 1.6$ | $10.6 \pm 2.2$ | $1.8 \pm 0.3$ | $264 \pm 44$ |
| 12 | $20.9 \pm 4.0$ | $13.7 \pm 4.8$ | $1.6 \pm 0.2$ | $233 \pm 26$ |
| 10 | $14.5 \pm 4.7$ | $15.8 \pm 4.1$ | $1.3 \pm 0.2$ | $185 \pm 26$ |
| 8 | $17.3 \pm 6.5$ | $17.8 \pm 4.2$ | $1.0 \pm 0.1$ | $148 \pm 19$ |

Table 6. Molecule Coverage (refers to number of alkanedithiols in a 2x2 simulation cell) Dependence of Thermal Conductivities and Thermal Conductance

From Table 6, no trend of in-plane thermal conductivities is found. The mean value of out-of-plane thermal conductivities/conductance decreases with decrease of coverage as expected. In Au-SAM-Au junctions, thermal energy is transported from one substrate to another through the discrete alkanedithiol molecules. These molecules are like channels through which energy passes. When the number of molecules decreases, energy transport channels are decreased. As a result, thermal energy transport becomes more difficult and conductance becomes smaller.

### 3.9. Vibration Coupling Analysis

To investigate how efficiently the thermal energy is transported from the substrate to the discrete molecules, the vibrational power spectra (also regarded as partial phonon density of state) of Au atoms in the substrate and sulfur atoms in the alkanedithiol molecules were calculated (see Figure 11). The calculations were done by performing Fourier transforms of the velocity autocorrelation functions. Two layers of Au atoms and 16 sulfur heads which form an interface were chosen as the samples from which the velocity autocorrelation functions were calculated. It can be seen that the vibration power spectrum of the Au substrate atoms has broader peaks than the sulfur, while the spectrum of the sulfur atoms has many discrete spikes. The overlap of these two spectra is limited to frequencies lower than 10THz. It is believed that the relatively large coupling is due to the strong Morse interaction between the Au and S atoms. Although the vibration coupling between Au and S appears to be strong, the discrete SAM molecules limit the number of channels available for thermal energy transport, and thus they significantly influence the heat transfer efficiency. As a result, the Au-SAM interface presents a large resistance to the junction. The calculated thermal conductance values in this work are close to those of the interface conductance values of water-surfactant heads ( $300 \pm 40$ $MW/(m^2K)$ ), hexane-surfactant tails ( $370 \pm 40$ $MW/(m^2K)$ ) and benzene-surfactant tails ( $200 \pm 30$ $MW/(m^2K)$ ) which also have large vibration coupling [7]. The C84-orgnanic solvent interface thermal conductance found by Huxtable et. al. [40] ( $10-20$ $MW/(m^2K)$ ) has almost no vibration spectra overlap, and is much lower than our data.

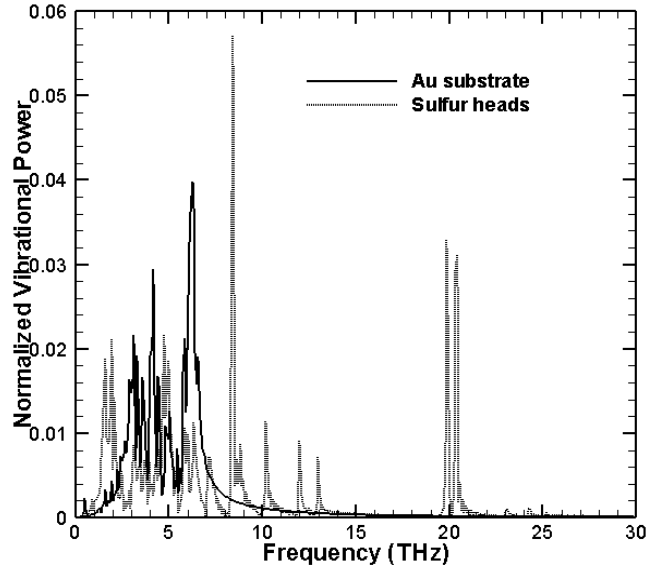

Figure 11. Normalized Vibrational Power Spectra of Au Substrate and Sulfur Heads

## 4. Summary and Conclusion

The present work calculated thermal conductivity/conductance of Au-SAM (alkanedithiol)-Au junctions using equilibrium classical MD with Green-Kubo method. PBCs were used in x- and y-directions. Both free boundary condition and PBC were used in the z-direction. The effect of simulation finite size was investigated. Vibration coupling was analyzed to explore the mechanism of thermal energy transport between Au substrates and SAM molecules. Due to the limited thermal transport channels presented by the discrete SAM molecules, thermal energy transport across the interface is not efficient even though the Au-S vibration coupling appears to be strong, and thus the interface resistance is large. From the phononic point of view, the Au-SAM interfaces present scattering sites which scatter and reflect the phonon wave packets. Only a part of the phonon energy is transmitted through the interfaces and this leads to interface resistance. Temperature dependence, molecule chain length dependence, substrate thickness dependence, pressure dependence and alkanedithiol molecule coverage dependence of thermal conductivities/conductance were studied. The results show that the thermal conductance in the out-of-plane direction (z-direction) increases with temperature increase at temperatures below 150K and remains almost unchanged at temperatures above 150K. The in-plane thermal conductivity displayed a bulk crystal lattice thermal conductivity behavior which decreases with the increase of temperature. It was also observed that the junction thermal conductance does not have obvious molecule chain length dependence (chain length from $-S-(CH_2)_8-S-$ to $-S-(CH_2)_{10}-S-$). Substrate thickness does not seem to affect the thermal conductance across the junction. Pressure dependence is also not obvious in Au-SAM-Au junctions. These three observations demonstrate that it is the Au-SAM interface that dominates the thermal transport across the junction. Alkanedithiol molecule coverage has an effect on the out-of-plane direction thermal transport. The thermal conductance decreases obviously with coverage decrease due to the reduced number of energy transport channels. All the calculated thermal conductance values are between $200$ and $300 MW/(m^2 K)$ which is inside the experimentally measured range of metal-nonmetal interfaces.

## Acknowledge

The authors gratefully acknowledge the support of NSF Grant Award ID 0522594 to enable this work to be performed. The project is in collaboration with Professor Arun Majumdar in UC Berkeley.

## References

[1]. Y. Loo, J. W. P. Hsu, R. L. Willett, K. W. Baldwin, K. W. West and J. A. Rogers. 2002. High-resolution transfer printing on GaAs surfaces using alkane dithiol monolayers. J. Vac. Sci. Technol. B, **20**, 2853

[2]. O. S. Nakagawa, S. Ashok, C. W. Sheen, J. Martensson and D. L. Allara. 1991. GaAs interface with Octadecyl Thiol Self-Assembled Monolayer: Structural and Electronical Properties. Jap. J. App. Phy. Vol. **30**, pp. 3759-3762

[3]. A. Koike and M. Yoneya. 1996. Molecular dynamics simulations of sliding friction of Langmuir-Blodgett mnolayers. J. Chem. Phys. **105**, 6060

[4]. C. L. McGuiness, A. Shaporenko, C. K. Mars, S. Uppili, M. Zharnikov and D. L. Allara. 2006. Molecular Self-Assembly at Bare Semiconductor Surfaces: Prparation and Characterization of Highly Organized Octadecanethiolate Monolayers on GaAs(001). J. Am. Chem. Soc. **128**, 5231-5243

[5]. Z. Ge, D. G. Cahill, and P. V. Braum. 2006. Thermal Conductance of Hydrophilic and Hydrophobic Interfaces. Phys. Rev. Let. **96**, 186101

[6]. Z. Wang, J. A. Carter, A. Lagutchev, Y. K. Koh, N. H. Seong, D. G. Cahill, and D. D. Dlott. 2007. Ultrafast Flash Thermal Conductance of Molecular Chains. Science, Vol. **317**. no. 5839, pp. 787 – 790

[7]. H. A. Patel, S. Garde, and P. Keblinski. 2005. Thermal Resistance of Nanoscopic Liquid-Liquid Interfaces: Dependence on Chemistry and Molecular Architecture. Nano Letters. **5**, 2225

[8]. Y. Yourdshahyan, H. K. Zhang, and A. M. Rappe. 2001. n-alkyl thiol head-group interactions with the Au(111) surface. Phys. Rev. B. **63**, 081405

[9]. H. Gronbeck, A. Curioni, and W. Andreoni. 2000. Thiols and Disulfides on the Au(111) Surface: The Headgroup-Gold Interaction. J. Am. Chem. Soc. **122**, 3839

[10]. C. W. Sheen, J. Shi, J. Martensson, A. N. Parikh, and D. L. Allara. 1992. A New Class of Organized Self-Assembled Monolayers: Alkane Thiols on GaAs(100). J. Am. Chem. Soc. **114**, 1514

[11]. W. Andersoni, A. Curioni, H. Gronbeck. 2000, Density Functional Theory Approach to Thiols and Disulfides on Gold: Au(111) Surface and Clusters. International Journal of Quantum Chemistry. **80**, 598

[12]. M. D. Muhlbaier. 2005. Self-Assembled Monolayers for Nanofabrication and Nano-Scale Electronics.

[13]. J. Hautman and M. L. Klein. 1989. Simulation of a monolayer of alkyl thiol chains. J. Chem. Phys. **91**, 4994

[14]. D. G. Cahill, W. K. Ford, K. E. Goodson. 2003. Nanoscale thermal transport. Journal of Applied Physics. **93** (2), 793.

[15]. J. M. Ziman. 1960. Electrons and Phonons. Oxford University Press, New York

[16]. S. Volz, J. B. Saulnier, M. Lallemand, B. Perrin, P. Depondt and M. Mareschal. 1996. Transient Fourier-law deviation by molecular dynamics in solid argon. Phys. Rev. B, **54**, 340

[17]. R. H. H. Poetzsch and H. Böttger. 1994. Interplay of disorder and anharmonicity in heat conduction: Molecular dynamics study. Phys. Rev. B. **50**, 15757

[18]. J. P. Crocombette, G. Dumazer, and N. Q. Hoang. 2007. Molecular dynamics modeling of the thermal conductivity of irradiated SiC as a function of cascade overlap. J. App. Phys. **101**, 023527

[19]. A. J. H. McGaughey, M. Kaviany. 2004. Thermal conductivity decomposition and analysis using molecular dynamics simulations, Part II. Complex silica structures. International Journal of Heat and Mass Transfer. **47**, 1799

[20]. J. Li, L. Porter, S. Yip. 1998. Atomistic modeling of finite-temperature properties of crystalline $\beta-SiC$, II. Thermal conductivity and effects of point defects. Journal of Nuclear Materials. **255**, 139

[21] R. J. Stevens, P. M. Norris and L. V. Zhigilei, 2004, Molecular-Dyanmics Study of Thermal Boundary Resistance: Evidence of Strong Inelastic Scattering Transport Channels, Proceeding of IMECE04.

[22]. R. Kubo, M. Yokota and S. Nakajima, 1957, Statistical-Mechanical Theory of Irreversible Processes. II. Response to Thermal Disturbance, J. Phys. Soc. Jpn. **12**, 1203.

[23]. C. L. Tien, J. R. Lukes and F. C. Chou, 1998, Molecular dynamics simulation of thermal transport in solids, Micro. Thermo. Eng., **2**, pp133-137.

[24]. B. L. Huang, A. J. H. McGaughey, and M. Kaviany. 2006. Thermal conductivity of metal-organic framework 5 (MOF-5): Part I. Molecular dynamics simulations. Int. J. Heat and Mass Transfer. **50**, pp. 405-411

[25]. A. R. Leach. 1996. Molecular Modeling Principles and Applications. Addison Wesley Longman Ltd. MA.

[26]. I. H. Sung, D. E. Kim. 2005. Molecular dynamics simulation study of the nano-wear characteristics of alkanethiol self-assembled monolayers. Appl. Phys. A. **81**, 109

[27]. R. Mahaffy, R. Bhatia, and B. J. Garrison. 1997. Diffusion of a Butanethiolate Molecule on a Au(111) Surface. J. Phys. Chem. B. **101**, 771

[28]. R. C. Lincoln, K. M. Koliwad, and P. B. Ghate. 1967. Morse-Potential Evaluation of Second- and Third-Order Elastic Constants of Some Cubic Metals. The Physical Review. **157**. 463

[29]. W. L. Jorgenson. 1989. Intermolecular potential functions and Monte Carlo simulations for liquid sulfur compounds. J. Phys. Chem. **90**, 6379-6388

[30]. A. K. Rappe, C. J. Casewit, K. S. Colwell, W. A. Goddard, and W. M. Skiff. 1992. UFF, a

Full Periodic Table Force Field for Molecular Mechanics and Molecular Dynamics Simulations. J. Am. Chem. Soc. **114**, 10024

[31]. L. Zhang, W. A. Goodard, S. Jiang. 2002. Molecular simulation study of the c(4x2) supperlattice structure of alkanethiol self-assembled monolayers on Au(111). J. Chem. Phys. **117**, 7342

[32]. H. Kaburaki, J. Li, S. Yip. 1999. thermal conductivity of solid argon by classical molecular dynamics. Mater. Res. Soc. Symp. Proc. **538**, 503

[33]. R. Y. Wang, R. A. Segalman, A. Majumdar. 2006. Room temperature thermal conductance of alkanedithiol self-assembled monolayers. Appl. Phys. Lett. **89**, 173113

[34]. Private discussion with Robert Wang from Mechanical Engineering of UC Berkeley.

[35]. R. J. Stoner and H. J. Maris. 1993. Kapitza conductance and heat flow between solids at temperatures from 50 to 300 K. Physical Review B. **48**, 16373

[36]. R. M. Costescu, M. A. Wall, and D. G. Cahill. 2003. Thermal conductance of epitaxial interfaces. Physical Review B. **67,** 054302

[37]. A. N. Smith, J. L. Hostetler, and P. M. Norris,.2000. Thermal boundary resistane measurements using a transient thermoreflectance technique. Microscale Thermophysical Engineering. **4**, 51

[38]. C. J. Gomes, M. Madrid, J. V. Goicochea, C. H. Amon. 2006. In-Plane and out-of-plane thermal conductivity of silicon thin films predicted by molecular dyanmcs. Transactions of the ASME. **128**, 1114

[39]. R. C. Lincoln, K. M. Koliwad, and P. B. Ghate. 1967. Morse-Potential Evaluation of Second- and Third-order Elastic Constants of Some Cubic Metals. The Physical Review. **157**, 463

[40]. S. T. Huxtable, D. G. Cahill, S. Shenogin, P. Keblinski, 2005, Relaxation of vibrational energy in fullerene suspensions, Chem. Phys. Lett., 407, 129-134